\documentclass[]{JHEP3} 

\usepackage{epsfig,multicol,bbm}
\usepackage[sort&compress,numbers]{natbib}

\newcommand\fverb{\setbox\fverbbox=\hbox\bgroup\verb}
\newcommand\fverbdo{\egroup\medskip\noindent%
			\fbox{\unhbox\fverbbox}\ }
\newcommand\fverbit{\egroup\item[\fbox{\unhbox\fverbbox}]}
\newbox\fverbbox

\usepackage{graphicx}
\def\tr{\mathop{\rm tr}\nolimits}
\def\Lsnew{\mathop{\rm \widetilde {Ls}}\nolimits}
\def\phid{\phi^\dagger}
\def\wh{\widehat}
\def\Ls{\mathrm{Ls}}
\def\Li{\mathrm{Li}}
\def\Ll{\mathrm{L}}
\def\I33m{\mathrm{I}_3^{3{\mathrm m}}}
\def\nn{\nonumber}
\def\fl{{\rm f}}
\def\be{\begin{equation}}
\def\ee{\end{equation}}
\def\bea{\begin{eqnarray}}
\def\eea{\end{eqnarray}}
\def\treenum{{(0)}}
\def\oneloopnum{{(1)}}
\def\nloopnum{{(n)}}
\def\qb{{\bar{q}}}
\def\cg{c_\Gamma}
\def\e{\epsilon}

\def\eps{\epsilon}

\def\spa#1.#2{\left\langle#1\,#2\right\rangle}
\def\spb#1.#2{\left[#1\,#2\right]}
\def\lor#1.#2{\left(#1\,#2\right)}
\def\sand#1.#2.#3{%
\left\langle\smash{#1}{\vphantom1}^{-}\right|{#2}%
\left|\smash{#3}{\vphantom1}^{-}\right\rangle}
\def\sandp#1.#2.#3{%
\left\langle\smash{#1}{\vphantom1}^{-}\right|{#2}%
\left|\smash{#3}{\vphantom1}^{+}\right\rangle}
\def\sandpp#1.#2.#3{%
\left\langle\smash{#1}{\vphantom1}^{+}\right|{#2}%
\left|\smash{#3}{\vphantom1}^{+}\right\rangle}
\def\sandpm#1.#2.#3{%
\left\langle\smash{#1}{\vphantom1}^{+}\right|{#2}%
\left|\smash{#3}{\vphantom1}^{-}\right\rangle}
\def\sandmp#1.#2.#3{%
\left\langle\smash{#1}{\vphantom1}^{-}\right|{#2}%
\left|\smash{#3}{\vphantom1}^{+}\right\rangle}
\def\spab#1.#2.#3{\langle#1|#2|#3]}
\def\spba#1.#2.#3{[#1|#2|#3\rangle}
\def\g#1{#1_g}  

\title{Analytic results for the one-loop NMHV $H\bar{q}qgg$ amplitude}

\author{Simon Badger$^\dag$,
    \ John M. Campbell$^*$,
    \ R. Keith Ellis$^\ddag$,
    \ Ciaran Williams$^\flat$
    \\
    $^\dag$Deutches Elektronen-Synchrotron DESY, Platanenallee, 6, D-15738 Zeuthen, Germany
    \\
    $^*$Department of Physics and Astronomy, University of Glasgow, Glasgow, G12 8QQ,  UK
    \\
    $^\ddag$Fermilab, Batavia, IL 60510, USA
    \\
    $^\flat$Department of Physics, University of Durham, Durham, DH1 3LE, UK
    \\
    E-mails: 
    {\tt simon.badger@desy.de},
    {\tt j.campbell@physics.gla.ac.uk}, 
    {\tt ellis@fnal.gov}, 
    {\tt ciaran.williams@durham.ac.uk}.}

\received{\today} 		
\accepted{\today}		

\preprint{
DESY 09-180 \\
FERMILAB-PUB-09-505-T \\
IPPP/09/86}

\abstract{ We compute the one-loop amplitude for a Higgs boson, a quark-antiquark pair and a pair of gluons of
negative helicity, i.e. for the next-to-maximally helicity violating (NMHV) case,
${\cal A}(H,1_{\bar{q}}^-,2_q^+, \g3^-,\g4^-)$. The calculation is
performed using an effective Lagrangian which is  valid in the limit of very large top quark mass. As a result of
this paper all amplitudes for the transition of a Higgs boson into 4 partons are now known analytically at one-loop
order. }


\begin{document} 


\section{Introduction}
The hunt for the standard model Higgs boson is about to enter the endgame phase.
The lower and upper limits coming from direct searches at LEP~\cite{Barate:2003sz}
and indirect constraints from precision electroweak data from the Tevatron and
LEP~\cite{EWWG} are now supplemented by the first  
direct limit from a hadron collider~\cite{:2009pt,Krumnack:2009nz}.
With the increase of luminosity at the Tevatron 
and the advent of running at the LHC, a discovery or a more stringent set of limits 
is to be expected.

An important search channel for the Higgs boson, in the mass range $115 < m_H < 160$~GeV,
is production via weak boson fusion~\cite{Djouadi:2005gi}.
A Higgs boson produced in this channel
is expected to be produced relatively centrally, in association with two
hard forward jets. These striking kinematic features are expected
to enable a search for such events despite the otherwise overwhelming QCD backgrounds.
Confidence in the theoretical prediction for the Higgs signal process is based upon
knowledge of next-to-leading order corrections in both QCD~\cite{Han:1992hr,Figy:2003nv,Berger:2004pc}
and in the electroweak sector~\cite{Ciccolini:2007jr,Ciccolini:2007ec}.

However, in addition to the weak process, a significant number of such events may also
be produced via the strong interaction. In order to accurately predict the signal and,
in particular, to simulate faithfully the expected significance in a given Higgs model,
a fully differential NLO calculation of QCD production of a Higgs and two hard jets is
also required.

In the Standard Model the Higgs couples to two gluons via a top-quark loop. Calculations which involve
the full dependence on $m_t$ are difficult and a drastic simplification can be achieved if one works in an
effective theory in which the mass of the top quark is large~\cite{Wilczek:1977zn,Djouadi:1991tka,Dawson:1990zj}.
For inclusive Higgs production this approximation is valid over a wide range of Higgs masses and, for processes with
additional jets, the approximation is justified provided that the transverse
momentum of each jet is smaller than $m_t$~\cite{DelDuca:H2j1}. Tree-level calculations have been performed in both the
large-$m_t$ limit~\cite{Dawson:Htomultijet,Kauffman:H2jets} and with the exact-$m_t$
\cite{DelDuca:H2j1} dependence. 

Results for the one-loop corrections to all of the Higgs + 4 parton processes have been published
in 2005~\cite{Ellis:2005qe}. Although analytic results were provided for the
Higgs $\bar{q}q \bar{q}q $ processes, the bulk of this calculation was performed
using a semi-numerical method. In this approach the loop integrals were calculated
analytically whereas the coefficients with which they appear in the
loop amplitudes were computed numerically using a recursive method. 
Although some phenomenology was performed using this calculation~\cite{Campbell:2006xx},
the implementation of fully analytic formulae will lead to a faster 
code and permit more extensive phenomenological investigations. 

In recent years enormous progress has been made in solving the
problem of evaluating virtual corrections to NLO scattering
processes. Building upon the remarkable work of Bern, Dixon, Dunbar
and Kosower during the mid-nineties~\cite{Bern:1994zx,Bern:1994cg},
unitarity constructions for these virtual corrections have developed
into an efficient algebraic technique. The modern
generalised unitarity method utilises quadruple cuts with complex
momenta to freeze four dimensional loop momenta and uniquely determine
the box coefficients~\cite{Britto:2004nc}.  The computation of
triangle and bubble coefficients is also reduced to an algebraic
procedure by application of an OPP style integrand reduction
\cite{delAguila:2004nf,Ossola:2006us,Ellis:2007br} or using direct
analytic extraction~\cite{Forde:2007mi}.
Further developments employing D-dimensional cutting techniques
\cite{Bern:1995db,Bern:1996ja,Anastasiou:2006jv,Giele:2008ve} extend
the method to compute full one-loop amplitudes.  The procedure is well
suited to numerical implementations and a number of automated
approaches have been developed to the point of phenomenological
applications
\cite{Berger:2008sj,Berger:2008sz,Berger:2009zg,Berger:2009ep,
Ellis:2009zw,KeithEllis:2009bu,Ossola:2007ax,Bevilacqua:2009zn,Melnikov:2009dn}.

In this paper we derive a compact analytic formula for the Higgs NMHV amplitude
with a quark-antiquark pair and two like-helicity gluons. This is achieved by splitting the real Higgs scalar into two
complex scalars ($\phi$ and $\phi^{\dagger}$) such that the Higgs amplitude is recovered in the sum~\cite{Dixon:2004za}.
Coefficients of box, triangle and bubble integrals are computed by applying the generalised
unitarity method in four-dimensions. The rational terms are extracted from a Feynman diagram
computation which is simplified using the knowledge of unphysical
singularities in the cut-constructible terms.

The paper is organised
as follows. In Section \ref{sec:efflag} we describe the large top-mass
approximation and the decomposition of the Higgs into self-dual
($\phi$) and anti-self-dual ($\phi^\dagger$) components. 
Section \ref{sec:colour} 
recalls the colour decomposition into primitive amplitudes and 
section \ref{sec:review} provides a guide to the current literature on Higgs + 4 parton amplitudes
and recalls the known analytic results for $\phi \bar q qgg$ amplitudes that are needed
to construct results for the Higgs boson.
In section \ref{sec:results}
we present analytic results that are sufficient
for a complete description of the NMHV amplitude. We numerically evaluate the obtained expressions
for the one-loop colour ordered amplitudes in section
\ref{sec:numerical} before drawing our conclusions.

\section{Effective Lagrangian \label{sec:efflag}}
Our calculation is performed using an effective Lagrangian to express the coupling of gluons to 
the Higgs field~\cite{Wilczek:1977zn},
\be \label{EffLag}
\mathcal{L}_H^{\mathrm{int}} = \frac{C}{2} \, H\,\tr
G_{\mu\nu}\,G^{\mu\nu}\, . \label{Lint} \ee
This Lagrangian is obtained by replacing the full one-loop coupling of the Higgs boson to the gluons
via an intermediate top quark loop, by an effective local operator. The effective Lagrangian 
approximation is valid  in the limit $m_H < 2 m_t$. 
At the order required in this paper, the coefficient $C$  
is given by~\cite{Djouadi:1991tka,Dawson:1990zj},
\be
C =\frac{\alpha_S}{6 \pi v} \Big( 1 +\frac{11}{4 \pi} \alpha_S\Big)
 + {\cal O}(\alpha_S^3) \;.
\ee
Here $v$ is the vacuum expectation value of the Higgs field, $v = 246$ GeV. 
The trace in Eq.~(\ref{EffLag}) is over the colour degrees of freedom
which, since SU(3) generators in the fundamental representation are normalised 
such that $\tr T^a T^b = \delta^{ab}$, implies that
$\tr G_{\mu\nu}\,G^{\mu\nu} = G^a_{\mu\nu}\,G^{a\,\mu\nu}$.

Following reference~\cite{Dixon:2004za} we will introduce a complex scalar field,
\be 
\phi = \frac{1}{2} \left( H+i A \right),\;\;\;\; \phid = \frac{1}{2} \left( H-iA \right) \;,
\ee
so that the effective Lagrangian, Eq.~(\ref{EffLag}), can be written as,
\begin{eqnarray}
\mathcal{L}_{H,A}^{\mathrm{int}} 
&=& \frac{C}{2} \Big [H\,\tr G_{\mu\nu}\,G^{\mu\nu}+i A\,\tr G_{\mu\nu}\,{}^*G^{\mu\nu}\Big] \nn \\
&=&  C \Big [\phi\,\tr G_{\scriptscriptstyle{SD}\; \mu\nu}\,G^{\mu\nu}_{\scriptscriptstyle{SD}}
            +\phid\,\tr G_{\scriptscriptstyle{ASD}\; \mu\nu}\,G^{\mu\nu}_{\scriptscriptstyle{ASD}} \Big]  \;,
\end{eqnarray}
where the gluon field strength has been separated into a self-dual and an anti-self-dual component,
\be G_{\scriptscriptstyle{SD}}^{\mu\nu} = \frac{1}{2}
(G^{\mu\nu}+{}^*G^{\mu\nu})\, , \quad
G_{\scriptscriptstyle{ASD}}^{\mu\nu} = \frac{1}{2}
(G^{\mu\nu}-{}^*G^{\mu\nu})\, , \quad {}^*G^{\mu\nu} \equiv
\frac{i}{2} \e^{\mu\nu\rho\sigma} G_{\rho\sigma}\, .
\ee
Calculations performed in terms of the field $\phi$ are simpler 
than the calculations for the Higgs boson and, moreover, the amplitudes for $\phid$
can be obtained by parity.
In the final stage, the full Higgs boson amplitudes are then written as a combination of  $\phi$ and $\phid$ components:
\bea
A(H,\{p_k\})&=&A(\phi,\{p_k\})+A(\phid,\{p_k\}) \;, \nn \\
A(A,\{p_k\})&=& -i \left( A(\phi,\{p_k\})-A(\phid,\{p_k\}) \right) \;.
\eea

\section{Definition of colour ordered amplitudes \label{sec:colour}}
The colour decomposition of the $H\bar{q}qgg$ amplitudes is exactly the same as for the
case $\bar{q}qgg$ which was written down in ref.~\cite{Bern:1994fz}. For the tree graph
there are two colour stripped amplitudes,
\be
{\cal A}_4^\treenum(\phi,1_\qb,2_q,\g3,\g4)
 = Cg^2 \, \sum_{\sigma\in S_2}\left(
T^{a_{\sigma(3)}}T^{a_{\sigma(4)}}\right)_{i_2}^{\,\,\,\bar{\imath}_1}
A_4^\treenum(\phi,1_\qb,2_q,\sigma(3),\sigma(4)) \,.
\label{qqggtreedecomp}
\ee
At one-loop level the colour decomposition is, 
\bea
{\cal A}_4^\oneloopnum(\phi,1_\qb,2_q,\g3,\g4) &=& Cg^4\, \cg
\Bigg[ N_c\sum_{\sigma\in S_2}\left(
T^{a_{\sigma(3)}}T^{a_{\sigma(4)}}\right)_{i_2}^{\,\,\,\bar{\imath}_1}
A_{4;1}(\phi,1_\qb,2_q,\sigma(3),\sigma(4))
\nonumber\\ && \hskip1.3cm \null
+ \delta^{a_3 a_4} \, \delta_{i_2}^{\,\,\,\bar{\imath}_1}
A_{4;3}(\phi,1_\qb,2_q;\g3,\g4) \Bigg] \,.
\label{qqggloopdecomp}
\eea
In these equations $g$ is the strong coupling constant and
$\cg$ is the ubiquitous one-loop factor,
\be
\cg \equiv  {1\over(4\pi)^{2-\eps}}
   {\Gamma(1+\eps)\Gamma^2(1-\eps)\over\Gamma(1-2\eps)} \,.
\label{cgdefn}
\ee
The colour stripped amplitudes $A_{4;1}$ and $A_{4;3}$ 
can further be decomposed into primitive amplitudes,
\bea A_{4;1}(\phi,1_\qb,2_q,\g3,\g4) &=& A_4^L(\phi,1_\qb,2_q,\g3,\g4) -
\frac{1}{N_c^2}A_4^R(\phi,1_\qb,2_q,\g3,\g4) \nonumber\\ && \hskip0.0cm
\null + \frac{n_f}{N_c} \, A_4^\fl(\phi,1_\qb,2_q,\g3,\g4) \,,
\label{A41defn}
\eea
and,
\bea A_{4;3}(\phi,1_\qb,2_q;\g3,\g4) &=& 
   A_4^L(\phi,1_\qb,2_q,\g3,\g4) + A_4^R(\phi,1_\qb,2_q,\g3,\g4) + A_4^L(\phi,1_\qb,\g3,2_q,\g4)  \nonumber\\
&+&A_4^L(\phi,1_\qb,2_q,\g4,\g3) + A_4^R(\phi,1_\qb,2_q,\g4,\g3) + A_4^L(\phi,1_\qb,\g4,2_q,\g3)  \,.
\nonumber\\
\label{A43defn}
\eea
All of these colour decomposition equations, namely 
Eqs.~(\ref{qqggtreedecomp}, \ref{qqggloopdecomp}, \ref{A41defn}, \ref{A43defn})
are equally valid if the $\phi$ is replaced by a $\phid$ or a Higgs boson $H$.
Sample diagrams contributing to each of the primitive amplitudes
are shown in Figure~\ref{primitive}.
\FIGURE[ht]{
\parbox{10cm}{\center
\epsfig{file=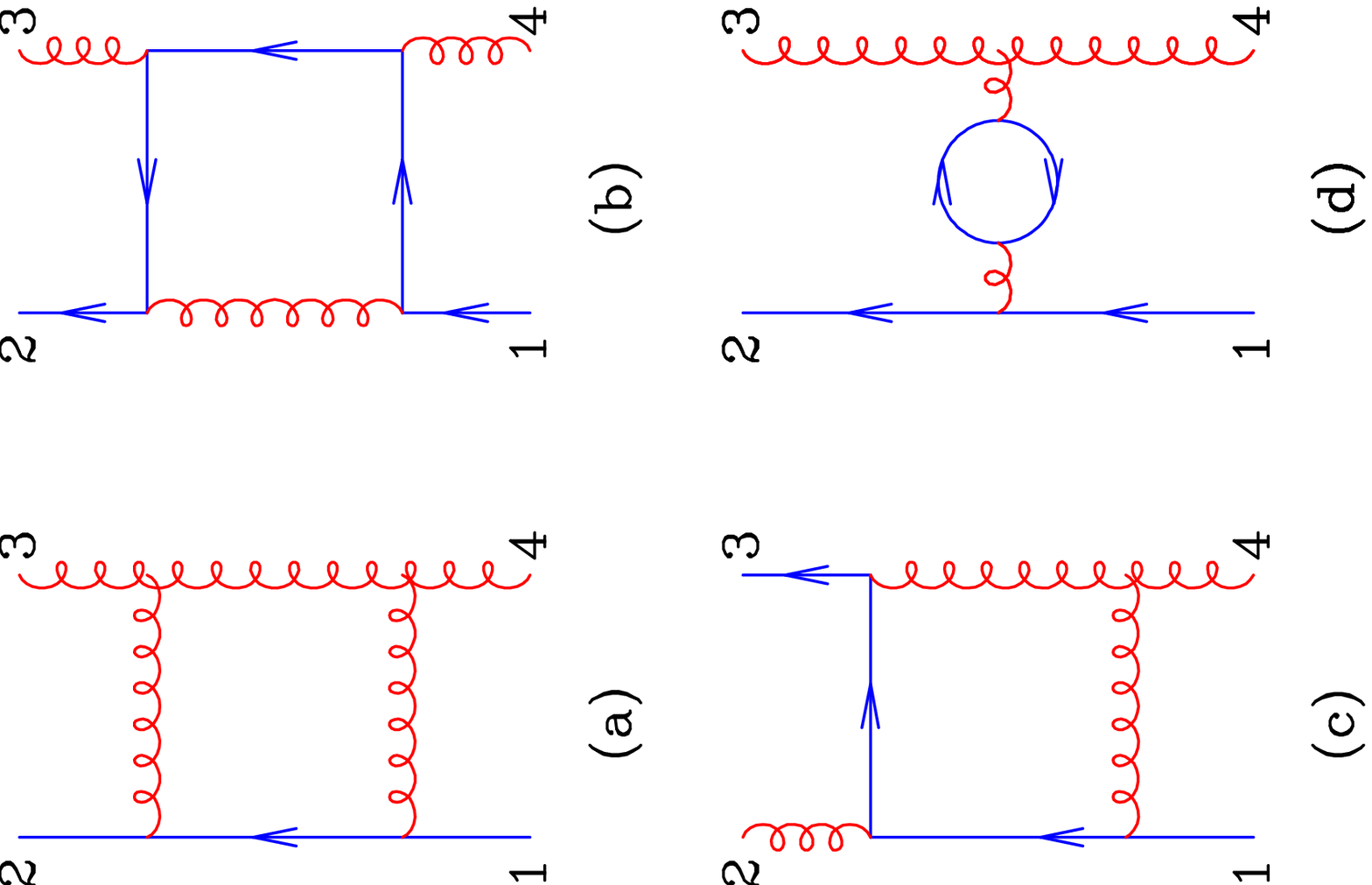,width=8cm,angle=270}}
\caption{Sample diagrams contributing to the primitive amplitudes 
(a) $A_4^L(\phi,1_\qb,2_q,\g3,\g4)$,
(b) $A_4^R(\phi,1_\qb,2_q,\g3,\g4)$,
(c) $A_4^L(\phi,1_\qb,\g2,3_q,\g4)$, and 
(d) $A_4^{\fl}(\phi,1_\qb,2_q,\g3,\g4)$.
The $\phi$ field can attach to any gluon line in the diagram.
\label{primitive}}
}

\section{Known analytic results for Higgs + 4 parton amplitudes \label{sec:review}}
In this section we review results from the literature and collect formulae,
for both tree and one-loop results, that will be useful in constructing the Higgs NMHV amplitude.

\subsection{Tree graph results \label{treeresults}}
The results for the tree graphs that are primarily of interest here, i.e. $\phi \bar q qgg$ amplitudes
with gluons of the same helicity, are:
\bea
- i A_4^\treenum(\phi,1_\qb^-,2_q^+,\g3^-,\g4^-)
&=&
- { {\spab3.{(1+4)}.2}^2 \spa4.1 \over \spb2.4 \, s_{124} }
    \biggl[ {1\over s_{12}} + {1 \over s_{41}} \biggr]
\nonumber\\
&& \hskip0.0cm \null
-  { {\spab4.{(1+3)}.2}^2 \spa1.3 \over \spb2.3 \, s_{12} \, s_{123} }
+  { {\spab1.{(3+4)}.2}^2 \over \spa1.2\spb2.4\spb2.3\spb3.4 }
\,, \label{phiaqggmpmmtree}\\
-i A_4^\treenum(\phi,1_\qb^-,2_q^+,\g3^+,\g4^+) &=& 0
\,, \label{phiaqggmppptree}
\eea
and for the subleading colour piece,
\bea
-i A_4^\treenum(\phi,1_\qb^-,\g2^-,3_q^+,\g4^-)
&=&
-  \, { {\spab4.{(1+2)}.3}^2 \over \spb1.2\spb2.3 \, s_{123} }
-  \, { {\spab2.{(1+4)}.3}^2 \over \spb3.4\spb4.1 \, s_{341} } \,,
\label{phiagqgmmpmtree}\\
-i A_4^\treenum(\phi,1_\qb^-,\g2^+,3_q^+,\g4^+) &=& 0\, .
\label{phiagqgmppptree}
\eea
A brief summary of our spinor notation is given in Appendix~\ref{spinnotation}.
Compact analytic expressions for all helicity amplitudes are presented
in references~\cite{Dixon:2009uk,Badger:2009hw}.

By using parity and charge conjugation~\cite{DelDuca:2004wt}, we can relate these $\phi \bar q qgg$ amplitudes
to ones for $\phid \bar q q gg$ with the same helicity assignments of quark and antiquark. This relation,
valid at any order of perturbation theory, $n$, reads,
\be
\mathcal{A}_4^\nloopnum(\phid,1_{\bar q}^{-h_q},2_q^{h_q},\g3^{h_3},\g4^{h_4})
= - \Big[
\mathcal{A}_4^\nloopnum(\phi,2_{\bar q}^{-h_q},1_q^{h_q},\g4^{-h_4},\g3^{-h_3})
\Big] \bigg|_{\spa{i}.{j} \leftrightarrow \spb{j}.{i} } \, .
\label{phiphidaggerparity}
\ee
We thus see that the $\phid$ amplitude in which we are interested is zero,
\be
-i A_4^\treenum(\phid,1_\qb^-,2_q^+,\g3^-,\g4^-) = 0 \;,
\label{phidaqggmpmmtree}
\ee
so that, at tree graph level, the NMHV Higgs amplitude in which we will ultimately be interested 
is simply given by Eq.~(\ref{phiaqggmpmmtree}).

\subsection{One-loop results}

We begin this section with a brief survey of the literature, to indicate where original calculations
of $H\bar{q}q\bar{q}q$ and $Hgggg$ amplitudes may be found. We then turn to the $H\bar{q}qgg$ amplitudes,
concluding the section by quoting the results for the $\phi^\dagger$ amplitude that must be combined
with the new $\phi$ amplitude calculation that we present here.

\subsubsection{$H\bar{q}q\bar{q}q$ amplitudes}

The full one-loop results for this process, both for pairs of identical and non-identical quarks, 
are already available in the literature. The matrix element squared has been computed in
ref.~\cite{Ellis:2005qe}, with results for the amplitude presented in ref.~\cite{Dixon:2009uk}. 

\subsubsection{$Hgggg$ amplitudes}
In principle there are $16$ combinations of amplitudes, but this number
is reduced to four independent amplitudes by parity and cyclicity.
The references to the complete set of needed amplitudes are given in Table~\ref{Hggggrefs}.
\TABLE{
\begin{tabular}{|l|l|l|}
\hline
$H$ amplitude & $\phi$ amplitude & $\phid$ amplitude\\
\hline
${\cal A}(H,1_g^+,2_g^+,\g3^+,\g4^+)$ &${\cal A}(\phi,1_g^+,2_g^+,\g3^+,\g4^+)$~\cite{Berger:2006sh}  &${\cal A}(\phid,1_g^+,2_g^+,\g3^+,\g4^+)$~\cite{Badger:2006us} \\
${\cal A}(H,1_g^-,2_g^+,\g3^+,\g4^+)$ &${\cal A}(\phi,1_g^-,2_g^+,\g3^+,\g4^+)$~\cite{Berger:2006sh}  &${\cal A}(\phid,1_g^-,2_g^+,\g3^+,\g4^+)$~\cite{Badger:2009hw}  \\
${\cal A}(H,1_g^-,2_g^-,\g3^+,\g4^+)$ &${\cal A}(\phi,1_g^-,2_g^-,\g3^+,\g4^+)$~\cite{Badger:2007si}  &${\cal A}(\phid,1_g^-,2_g^-,\g3^+,\g4^+)$~\cite{Badger:2007si}  \\
${\cal A}(H,1_g^-,2_g^+,\g3^-,\g4^+)$ &${\cal A}(\phi,1_g^-,2_g^+,\g3^-,\g4^+)$~\cite{Glover:2008ffa} &${\cal A}(\phid,1_g^-,2_g^+,\g3^-,\g4^+)$~\cite{Glover:2008ffa}  \\
\hline
\end{tabular} 
\caption{$\phi$ and $\phid$ amplitudes needed to construct a given one-loop $Hgggg$ amplitude, together with the references where they can be obtained.
In all cases the $\phid$ amplitudes are constructed from the $\phi$ amplitudes given in the reference using the parity operation.
Results for all helicity combinations are also written, in uniform notation, in ref.~\cite{Badger:2009hw}.
}
\label{Hggggrefs}
}
In addition a nice summary of all the one-loop results for the Higgs + 4 gluon 
amplitudes is given in ref.~\cite{Badger:2009hw}.

\subsubsection{$H\bar{q}qgg$ amplitudes}
In principle there are 8 combinations of amplitudes, since helicity is conserved on the quark line,
but because of parity invariance only four Higgs amplitudes are independent.
The references to the amplitudes already calculated in the literature are given in Table~\ref{Haqggrefs}.
\TABLE{
\begin{tabular}{|l|l|l|}
\hline
$H$ amplitude & $\phi$ amplitude & $\phid$ amplitude\\
\hline
${\cal A}(H,1_{\bar{q}}^-,2_q^+,\g3^+,\g4^+)$
&${\cal A}(\phi,1_{\bar{q}}^-,2_q^+,\g3^+,\g4^+)$~\cite{Berger:2006sh,Dixon:2009uk}    &${\cal A}(\phid,1_{\bar{q}}^-,2_q^+,\g3^+,\g4^+)$  \\
${\cal A}(H,1_{\bar{q}}^-,2_q^+,\g3^-,\g4^-)$ 
&${\cal A}(\phi,1_{\bar{q}}^-,2_q^+,\g3^-,\g4^-)$ &${\cal A}(\phid,1_{\bar{q}}^-,2_q^+,\g3^-,\g4^-)$~\cite{Berger:2006sh,Dixon:2009uk}\\
${\cal A}(H,1_{\bar{q}}^-,2_q^+,\g3^+,\g4^-)$ 
&${\cal A}(\phi,1_{\bar{q}}^-,2_q^+,\g3^+,\g4^-)$~\cite{Dixon:2009uk} &${\cal A}(\phid,1_{\bar{q}}^-,2_q^+,\g3^+,\g4^-)$~\cite{Dixon:2009uk}  \\
${\cal A}(H,1_{\bar{q}}^-,2_q^+,\g3^-,\g4^+)$ 
&${\cal A}(\phi,1_{\bar{q}}^-,2_q^+,\g3^-,\g4^+)$~\cite{Dixon:2009uk} &${\cal A}(\phid,1_{\bar{q}}^-,2_q^+,\g3^-,\g4^+)$~\cite{Dixon:2009uk} \\
\hline
\end{tabular}
\caption{$\phi$ and $\phid$ amplitudes needed to construct a given
one-loop $H\bar{q}qgg$ amplitude, together with the references where
they can be obtained.  In all cases the $\phid$ amplitudes are
constructed from the $\phi$ amplitudes given in the reference, using
the parity operation.  The cases where the gluons have the same
helicity, which have no associated references, are the subject of this
paper.}\label{Haqggrefs}}
From this table we see that the Higgs amplitude ${\cal A}(H,1_{\bar{q}}^-,2_q^+,\g3^-,\g4^-)$  requires, in addition
to the calculation of a previously unknown $\phi$ amplitude, also the results for the corresponding $\phid$ amplitude
from ref.~\cite{Dixon:2009uk}.

The $\phid$ results that we shall need can be derived from the following amplitudes in the case of $A_{4;1}$,
\bea
&&\null -i A_4^L(\phi,1_\qb^-,2_q^+,\g3^+,\g4^+)\ =\
2 i \, A_4^\treenum(\phid,1_\qb^-,2_q^+,\g3^+,\g4^+)
\nonumber\\ && \hskip2.0cm \null
+ {1\over2} \biggl[
   { \spab1.{(2+3)}.4 \over \spa2.3 \spa3.4 }
   + { \spa1.2 \spb2.3 \spa3.1 \over \spa2.3 \spa3.4 \spa4.1 }
 \biggr] 
- {1\over3} { \spa1.3 \spb3.4 \spa4.1 \over \spa1.2 {\spa3.4}^2 } \,,
\label{qqgg_mppp_L} \\
&&\null -i A_4^R(\phi,1_\qb^-,2_q^+,\g3^+,\g4^+)\ =\
- {1\over2} \biggl[
   { \spab1.{(2+3)}.4 \over \spa2.3 \spa3.4 }
   + { \spa1.2 \spb2.3 \spa3.1 \over \spa2.3 \spa3.4 \spa4.1 }
 \biggr] \,,
\label{qqgg_mppp_R} \\
&&\null -i A_4^{\fl}(\phi,1_\qb^-,2_q^+,\g3^+,\g4^+)\ =\
{1\over3} { \spa1.3 \spb3.4 \spa4.1 \over \spa1.2 {\spa3.4}^2 } \,,
\label{qqgg_mppp_fl}
\eea
whilst the subleading partial amplitude $A_{4;3}$ also requires the results,
\bea
-i A_4^L(\phi,1_\qb^-,\g2^+,3_q^+,\g4^+) &=&
- 2 \, { {\spab1.{(2+3)}.4}^2 \over \spa1.2 \spa2.3 \, s_{123} } 
+ {1\over2} \left[
{ \spa1.3 \spab1.{(3+4)}.2 \over \spa2.3\spa3.4\spa4.1 }
  + { {\spa1.3}^2 \spb3.4 \over \spa1.2\spa2.3\spa3.4 }
         \right]\,,
\nn\\ 
\label{phi_qgqg_mppp_L}
A_4^\fl(\phi,1_\qb^-,2^+,3_q^+,\g4^+) &=& 0.
\label{phi_qgqg_mppp_fl}
\eea

To obtain the form that is most useful for the calculation of
${\cal A}(H,1_{\bar{q}}^-,2_q^+,\g3^-,\g4^-)$,
we relate the $\phid \bar{q}qgg$ amplitudes to the $\phi\bar{q}qgg$ ones by using the
relation in Eq.~(\ref{phiphidaggerparity}).
Thus we obtain the required results by performing the transformation
$1 \leftrightarrow 2, 3 \leftrightarrow 4$,
$\left\langle \right\rangle \leftrightarrow \left[\right]$ and reversing the sign.
The amplitudes contributing to $A_{4;1}$ are,
\bea
&&\null -i A_4^L(\phid,1_\qb^-,2_q^+,\g3^-,\g4^-)\ =\
2 i \, A_4^\treenum(\phi,1_\qb^-,2_q^+,\g3^-,\g4^-)
\nonumber\\ && \hskip2.0cm \null
+{1\over2} \biggl[
   { \spab3.{(1+4)}.2 \over \spb1.4 \spb3.4 }
   + { \spb2.1 \spa1.4 \spb2.4 \over \spb1.4 \spb3.4 \spb2.3 } \biggr] 
-{1\over3} { \spb2.4 \spa3.4 \spb2.3 \over \spb1.2 {\spb3.4}^2 } \,.
\label{phidaqgg_mpmm_L}
\eea
\be
-i A_4^R(\phid,1_\qb^-,2_q^+,\g3^-,\g4^-)\ =\
 -{1\over2} \biggl[
   { \spab3.{(1+4)}.2 \over \spb1.4 \spb3.4 }
   + { \spb2.1 \spa1.4 \spb2.4 \over \spb1.4 \spb3.4 \spb2.3 }
 \biggr] \,,
\label{phidaqgg_mpmm_R}
\ee
\be
-i A_4^{\fl}(\phid,1_\qb^-,2_q^+,\g3^-,\g4^-)\ =\
{1\over3} { \spb2.4 \spa3.4 \spb2.3 \over \spb1.2 {\spb3.4}^2 } \,,
\label{phidaqgg_mpmm_fl}
\ee
while the additional subleading contributions become,
\bea
-i A_4^L(\phid,1_\qb^-,\g3^-,2_q^+,\g4^-) &=&
 2 \, { {\spab3.{(4+1)}.2}^2 \over \spb2.4 \spb4.1 \, s_{124} } 
-{1\over2} \left[
{ \spb2.1 \spab4.{(1+3)}.2 \over \spb4.1\spb1.3\spb3.2 }
  + { {\spb1.2}^2 \spa1.3 \over \spb2.4\spb4.1\spb1.3 }
         \right]
\nn\\
&=&  i \, A_4^\treenum(\phi,1_\qb^-,\g3^-,2_q^+,\g4^-)
 + \mbox{terms antisymmetric in}~ \{ 3\leftrightarrow 4 \} \;. \nn\\
\label{phig_qgqg_mmpm_L}
\eea
We note that all of these amplitudes are finite because of the vanishing
of the corresponding tree graph results (see section~\ref{treeresults}).

\section{One-loop results \label{sec:results}}

In this section we present analytic expressions for the full one-loop corrections to the process
$\mathcal{A}_{4}^{(1)}(\phi,1_\qb^-,2_q^+,\g3^-,\g4^-)$. All expressions are presented un-renormalised in the four-dimensional
helicity (FDH) scheme (setting $\delta_R=0$) or 't Hooft-Veltman scheme (setting $\delta_R=1$).

We employ the generalised unitarity method
\cite{Britto:2004nc,Forde:intcoeffs,Mastrolia:2006ki,Britto:sqcd,Britto:ccqcd} to calculate the
cut-constructible parts of the left-moving, right-moving and $n_f$ one-loop amplitudes. This
relies on the expansion of a one-loop amplitude in terms of scalar basis integrals, 
\bea
A^{\rm cut-cons.}_4(\phi,1_{\overline{q}}^-,2^+_q,\g3^-,\g4^-)=
\sum_{i}{C}_{4;i} \mathcal{I}_{4;i} + \sum_{i}{C}_{3;i} \mathcal{I}_{3;i} +\sum_{i}{C}_{2;i} \mathcal{I}_{2;i}.
\eea
In this sum each $j$-point scalar basis integral ($\mathcal{I}_{j;i}$) appears with a coefficient
$C_{j;i}$. The sum over $i$ represents the sum over the partitions of the external
momenta over the $j$ legs of the basis integral. 
Multiple cuts isolate different integral functions and allow the
construction of a linear system of equations from which the coefficients can be extracted. 
We use the quadruple cut method ~\cite{Britto:2004nc} which freezes the loop momenta and determines each 
box coefficient uniquely. Triangle coefficients are determined using the Laurent expansion method \cite{Forde:intcoeffs},
whilst the  two-point coefficients are determined {\it via} Stokes' Theorem applied to functions of 
two complex-conjugated variables \cite{Mastrolia:2009dr}.
Results were obtained using the QGRAF~\cite{Nogueira:1991ex}, FORM~\cite{Vermaseren:2008kw} and S@M~\cite{Maitre:2007jq}
packages in order to control the extensive algebra.

\subsection{Results for $A_{4;1}(\phi,1_\qb,2_q,\g3^-,\g4^-)$}
The partial amplitude $A_{4;1}(\phi,1_\qb,2_q,\g3^-,\g4^-)$ is calculated from three primitive
amplitudes according to Eq.~(\ref{A41defn}). We shall deal with each of these ingredients in turn.

\subsubsection{$A_4^L(\phi,1_\qb^-,2_q^+,\g3^-,\g4^-)$}
The full result for this primitive amplitude is given by,
\bea
&&\null -i A_4^L(\phi,1_\qb^-,2_q^+,\g3^-,\g4^-)\ =\
-i A_4^\treenum(\phi,1_\qb^-,2_q^+,\g3^-,\g4^-) \times V_1^L 
\nn\\
&&\null
+\frac{s_{134}^2}{\spb1.4 \spb3.4 \spab2.(1+4).3} \Big[ \Ls_{-1}(s_{14},s_{34};s_{134}) 
+\Lsnew^{2{\rm m}h}_{-1}(s_{12},s_{134};s_{34},m_\phi^2) \Big] \nn \\
&&\null
-\frac{{\spab1.(3+4).2}^2}{ {\spab1.(2+3).4} \spb2.3 \spb3.4} \Big[ \Ls_{-1}(s_{34},s_{23};s_{234}) 
+ \Lsnew^{2{\rm m}h}_{-1}(s_{12},s_{234};s_{34},m_\phi^2) \Big] \nn \\
\nn \\
&&\null
+\Big[ \frac{m_\phi^4 {\spa1.4}^2 \spa2.4}{\spa1.2 \spab2.(1+4).3 \spab4.(1+2).3 s_{124}}
 -\frac{{\spab3.(1+4).2}^3 }{ \spb1.2 \spb2.4 \spab3.(1+2).4 s_{124}}\Big]
\Ls_{-1}(s_{12},s_{14};s_{124}) \nn \\
&&\null
+\Big[\frac{{\spb2.3}^2 {\spab4.(2+3).1}^3}{\spb1.2 {\spb1.3}^3 \spab4.(1+2).3 s_{123}}
-\frac{m_\phi^4 {\spa1.3}^3}{\spa1.2 \spab1.(2+3).4 \spab3.(1+2).4 s_{123}}
\Big] \Ls_{-1}(s_{12},s_{23};s_{123}) \nn \\
&&\null
+\Big[ \frac{{\spab4.(1+3).2}^3}{\spb1.2 \spb2.3 \spab4.(1+2).3 s_{123}}
-\frac{m_\phi^4 {\spa1.3}^3}{\spa1.2 \spab1.(2+3).4 \spab3.(1+2).4 s_{123}} \Big] \nn \\ 
&&\null 
\times \Big[\Lsnew^{2{\rm m}h}_{-1}(s_{34},s_{123};s_{12},m_\phi^2) 
           +\Lsnew^{2{\rm m}h}_{-1}(s_{14},s_{123};s_{23},m_\phi^2) \Big] \nn \\
&&\null
+\Big[\frac{m_\phi^4 {\spa1.4}^2 \spa2.4}{\spa1.2 \spab2.(1+4).3 \spab4.(1+2).3 s_{124}}
-\frac{{\spab3.(1+4).2}^2  \spab3.(2+4).1 }{\spb1.2 \spb1.4 \spab3.(1+2).4 s_{124}} \Big] \nn \\
&&\null 
\times \Big[\Lsnew^{2{\rm m}h}_{-1}(s_{34},s_{124};s_{12},m_\phi^2) 
           +\Lsnew^{2{\rm m}h}_{-1}(s_{23},s_{124};s_{14},m_\phi^2) \Big] \nn \\
&&\null 
-C_{3;\phi|12|34}(\phi,1_\qb^-,2_q^+,\g3^-,\g4^-) \I33m(s_{12},s_{34},m_\phi^2)
-C_{3;\phi|41|23}(\phi,1_\qb^-,2_q^+,\g3^-,\g4^-) \I33m(s_{23},s_{14},m_\phi^2)
 \nn \\
&&\null
   -\frac{2 {\spa1.3}^2 \spa3.4 \spab4.(1+2).3 \spb1.2}{3} \wh{\Ll}_3(s_{123},s_{12}) \nn \\
&&\null
   +\frac{\spa3.4 \spa3.1
    (\spab4.(1+3).2 \spb1.3-3 \spab4.(2+3).1 \spb2.3 )}
    {6 \spb3.1}  \wh{\Ll}_2(s_{123},s_{12}) \nn \\
&&\null
    +\frac{\spa1.3
     \Bigl( 16 {\spab4.(1+3).2}^2 {\spb1.3}^2
    -3 \spab4.(1+3).2 \spab4.(1+2).3 \spb2.1 \spb3.1   
    +6 {\spab4.(2+3).1}^2 {\spb2.3}^2 \Bigr)}
    {6 s_{123} {\spb3.1}^2 \spb3.2} \nn \\
&& \quad \times \wh{\Ll}_1(s_{123},s_{12})
\nn \\
&&\null
    -\frac{2 s_{124} {\spa3.4}^2 \spa1.4 \spb4.2}{3} \wh{\Ll}_3(s_{124},s_{12})
\nn \\
&&\null
    -\spa3.4 \spa1.4
     \frac{2 \spab3.(1+4).2 \spb1.4 -3 \spab3.(1+2).4  \spb1.2}
    {6 \spb4.1 }  \wh{\Ll}_2(s_{124},s_{12})
\nn \\
&&\null
    -\frac{\spab3.(1+4).2  (9 \; s_{124} \spa3.4 \spb2.1
    -22 \; \spab3.(1+4).2 \spa4.2 \spb1.2)}
     {6 \; s_{124} \spb4.1 \spb2.1 } \wh{\Ll}_1(s_{124},s_{12})
\nn \\
&&\null
    +\frac{\spa1.4 \spa1.3 \spab4.(2+3).1 \spb1.2}
    {2 \spb3.1} \wh{\Ll}_2(s_{123},s_{23})
\nn \\
&&\null
    -\spa1.3 \spab4.(2+3).1
     \frac{3 \; \spab4.(1+3).2 \spb1.3 + 2 \; \spab4.(2+3).1 \spb2.3 }
    {2 s_{123} {\spb1.3}^2} \wh{\Ll}_1(s_{123},s_{23})
\nn \\
&&\null
    +\frac{s_{234} \spa1.4 \spa3.4 \spb4.2}
     {2 \spb4.3} \wh{\Ll}_2(s_{234},s_{23})
    +3 \frac{\spa3.4 \spab1.(3+4).2 }{2 \spb4.3 } \wh{\Ll}_1(s_{234},s_{23}) \nn \\
&&+R^{L}(\phi,1^-_{\overline{q}},2^{+}_q,\g3^-,\g4^-) \; ,
\label{qqgg_mpmm_L}
\eea
with,
\bea
V_1^L &=& - \frac{1}{\e^2} \left[
\left(\frac{\mu^2}{-s_{23}}\right)^\e +
\left(\frac{\mu^2}{-s_{34}}\right)^\e +
\left(\frac{\mu^2}{-s_{41}}\right)^\e\right] +
\frac{13}{6\e}\left(\frac{\mu^2}{-s_{12}}\right)^\e + \frac{119}{18}
- \frac{\delta_R}{6} \;,
\label{V_1_L}
\eea
and the remaining rational terms given by,
\bea
&&R^L(\phi,1_{\overline{q}}^-,2^+_q,\g3^-,\g4^-)= 
\frac{\spa3.4 \spab3.1+4.2 \Bigl( 2 \spa2.4 \spb4.2 - \spa1.2 \spb2.1 \Bigr)}{12s_{124}\spa1. 2\spb2. 1\spb4. 1}\nn\\&&
+\frac{\spa2. 3\spab4. (1+3). 2^2 \Bigl( 3\spa1. 2\spb2. 1 - 2\spa2. 3\spb3. 2 \Bigr) - 
   2\spa1. 3^2\spa2. 4\spab4. (2+3). 1\spb2. 1\spb3. 2}{
  12s_{123}\spa1. 2\spa2. 3\spb2. 1\spb3. 1\spb3. 2} \nn\\&&
+\frac{5\spa3. 4^2}{12\spa2. 3\spb3. 1}+ \frac{5\spa3. 4\spab4. (1+3). 2}{
  6\spa2. 3\spb3. 1\spb3. 2} + \frac{\spab4. (1+3). 2^2}{6\spa1. 2\spb2. 1\spb3. 1
   \spb3. 2}\nn\\&&
 - \frac{\spa1. 3\spa1. 4\spa2. 4\spb2. 1}{3\spa1. 2\spa2. 3\spb3. 1\spb3. 2} 
 - \frac{\spa1. 3\spa3. 4}{12\spa1. 2\spb4. 1} - 
 \frac{\spa3. 4^2\spb4. 2}{6\spa1. 2\spb2. 1\spb4. 1} + 
 \frac{\spa1. 3\spa2. 4\spab4. (1+3). 4}{4\spa1. 2\spa2. 3\spb3. 1\spb4. 3}\nn\\&& - 
 \frac{\spa1. 3\spab4. (1+3). 4}{3\spa1. 2\spb4. 1\spb4. 3} - 
 \frac{5\spa1. 4^2\spb4. 1}{12\spa1. 2\spb3. 1\spb4. 3} + 
 \frac{\spa1. 4^2\spb4. 2}{6\spa1. 2\spb3. 2\spb4. 3} \;.
\eea
The coefficients of the three mass triangles were calculated using the method of ref.~\cite{Forde:2007mi},
\be
C_{3;\phi|12|34}(\phi,1_\qb^-,2_q^+,\g3^-,\g4^-) = 
\sum_{\gamma=\gamma_{\pm}} \frac{ m_\phi^4 {\spa3.4}^3 {\spa 1.{K_1^{\flat}}}^2 }
{\gamma (\gamma-m_\phi^2) {\spa1.2} {\spa 3.{K_1^{\flat}}} {\spa 4.{K_1^{\flat}}}} \;,
\ee
with $K_1=-p_1-p_2-p_3-p_4,K_2=-p_1-p_2$
and the massless vector $K_1^\flat$ given by,
\be \label{K1flat}
K_1^{\flat\;\mu}= \gamma \frac{\gamma K_1^\mu -K_1^2 K_2^\mu}{\gamma^2-K_1^2 K_2^2} \;,
\ee
and where $\gamma$ is given by the two solutions,
\be
\gamma_{\pm} = K_1\cdot K_2 \pm \sqrt{(K_1 \cdot K_2)^2-K_1^2 K_2^2} \;.
\ee
The other triangle coefficient is,
\be
C_{3;\phi|41|23}(\phi,1_\qb^-,2_q^+,\g3^-,\g4^-) = -
\sum_{\gamma=\gamma_{\pm}} \frac{ m_\phi^4 {\spa1.4}^2 {\spa 3.{K_1^{\flat}}}^2 }
{2 \gamma (\gamma-m_\phi^2) {\spa 1.{K_1^{\flat}}} {\spa 2.{K_1^{\flat}}}} \;,
\ee
with $K_1=-p_1-p_2-p_3-p_4,K_2=-p_1-p_4$
and $K_1^\flat$ given in terms of these vectors by Eq.~(\ref{K1flat}).

The definitions of the box integral functions $\Ls_{-1}$ and $\Lsnew^{2{\rm m}h}_{-1}$
can be found in Appendix~\ref{intnotation}, together with expressions for
$\wh{\Ll}_1$, $\wh{\Ll}_2$ and $\wh{\Ll}_3$. In addition to logarithms and polynomial denominators, the latter
functions also contain rational terms that protect them from unphysical singularities. Thus, for example,
\be
\wh{\Ll}_2(s,t) = \frac{\log{(s/t)}}{(s-t)^2}-\frac{1}{2(s-t)} \left(\frac{1}{s}+\frac{1}{t}\right) \nn \; ,
\ee
which is finite in the limit that $s \to t$. 

\subsubsection{$A_4^R(\phi,1_\qb^-,2_q^+,\g3^-,\g4^-)$}
The result for the right-moving amplitude, $A_4^R(\phi,1_\qb^-,2_q^+,\g3^-,\g4^-)$ is,
\bea
&&-i A_4^R(\phi,1_\qb^-,2_q^+,\g3^-,\g4^-)\ =\
-i A_4^\treenum(\phi,1_\qb^-,2_q^+,\g3^-,\g4^-)\,\times V^R
\nn\\
&+&\frac{{\spb1.2}^2 {\spab4.(1+2).3}^2}
   {{\spb 1.3}^3 {\spb2.3} s_{123}}
   \; \Ls_{-1}(s_{12},s_{23};s_{123}) 
  +\frac{{\spab 3.(1+4).2}^2}
   {\spb 1.4 \spb 2.4 s_{124}}
  \; \Ls_{-1}(s_{14},s_{12};s_{124})  \nn \\
&-& \frac{{\spab1.(3+4).2}^2}
   {{\spb 2.3} {\spb 3.4} {\spab 1.(2+3).4}}
    \; \Lsnew^{2{\rm m}h}_{-1}(s_{14},s_{234};s_{23},m_\phi^2) \nn \\
&+& \frac{s_{134}^2}
   {{\spb 1.4}  {\spb3.4} {\spab 2.(1+4).3}}
   \; \Lsnew^{2{\rm m}h}_{-1}(s_{23},s_{134};s_{14},m_\phi^2) \nn \\
&-&
C_{3;\phi|41|23}(\phi,1_\qb^-,2_q^+,\g3^-,\g4^-) \I33m(s_{23},s_{14},m_\phi^2) 
  -\frac{1}{2} \frac{{\spa1.4}^2 {\spb1.2}^2 {\spab3.(1+2).4}^2}
  { \spb1.4 \spb2.4  s_{124}}
   \; \wh{\Ll}_2(s_{124},s_{12}) \nn \\
&+&2 \frac{\spa3.4  \spab3.(1+4).2}
  {\spb1.4}
    \; \wh{\Ll}_1(s_{124},s_{12})
  +\frac{1}{2} \frac{{\spab3.(1+4).2}^2}
   {\spb1.4  \spb2.4  s_{124}}
    \; \wh{\Ll}_0(s_{124},s_{12}) \nn \\
&-&
  \frac{1}{2} \frac{{\spa1.4}^2  {\spb2.4}^2  s_{234}^2}
   {\spb2.3 \spb3.4  \spab1.(2+3).4 }
   \; \wh{\Ll}_2(s_{234},s_{23})
  -2 \frac{\spa3.4 \spab1.(3+4).2 }
   {\spb3.4}
    \; \wh{\Ll}_1(s_{234},s_{23}) \nn \\
&+&
  \frac{1}{2} \frac{ {\spab1.(3+4).2}^2}
   { \spb2.3 \spb3.4  \spab1.(2+3).4 }
    \; \wh{\Ll}_0(s_{234},s_{23})
  -\frac{1}{2} \frac{ \Bigl( \spa1.2  \spb1.2  \spab4.(2+3).1 \Bigr)^2 \spb2.3}
   {{\spb1.3}^3 s_{123}}
    \; \wh{\Ll}_2(s_{123},s_{23}) \nn \\
&+&
  2 \frac{\spa1.3  \spb1.2  \spab4.(1+2).3  \spab4.(2+3).1 }
   {\spa2.3  {\spb1.3}^2 \spb2.3}
    \; \wh{\Ll}_1(s_{123},s_{23}) \nn \\
&+&
  \Big[-2 \frac{ \spa1.3  \spb1.2  \spab4.(1+2).3  \spab4.(2+3).1 }
   {s_{123} {\spb1.3}^2 \spa2.3 \spb2.3}
  +\frac{1}{2} \frac{{\spab4.(2+3).1}^2 \spb2.3 }{{\spb1.3}^3 s_{123}} \Big]
   \; \wh{\Ll}_0(s_{123},s_{23}) \nn \\
&-&
  \frac{1}{2} \frac{ \Bigl( \spa1.3 \spb1.2  \spab4.(1+2).3 \Bigr)^2} 
  { \spb1.3 \spb2.3  s_{123}}
   \; \wh{\Ll}_2(s_{123},s_{12}) \nn \\
&+&
  \spa3.4  \spb1.2  \spab4.(1+2).3
   \frac{(-2 \spa1.3 \spb1.3-\spa2.3 \spb2.3)}
  {\spa2.3 {\spb1.3}^2 \spb2.3}
   \; \wh{\Ll}_1(s_{123},s_{12}) \nn \\
&+&
  \spb1.2  \spab4.(1+2).3 
   \frac{ \spa2.3 \spab4.(1+3).2 + 2 \spa1.3 \spab4.(2+3).1 }
  {{\spb1.3}^2 \spa2.3 \spb2.3  s_{123}}
   \; \wh{\Ll}_0(s_{123},s_{12}) \nn \\
&+& R^{R}(\phi,1^-_{\overline{q}},2^{+}_q,\g3^-,\g4^-) \; ,
\label{qqgg_mpmm_R}
\eea
with
\be
V^R\ =\ - \frac{1}{\e^2}\left( \frac{\mu^2}{-s_{12}}\right)^\e
- \frac{3}{2\e}\left( \frac{\mu^2}{-s_{12}}\right)^\e
- \frac{7}{2} - \frac{\delta_R}{2}
\,.
\ee
The remaining rational pieces in Eq.~(\ref{qqgg_mpmm_R}) have the following form:
\bea
R^{R}(\phi,1^-_{\overline{q}},2^{+}_q,\g3^-,\g4^-) =
- \frac{\spa2. 4^2\spb2. 1^2}{2\spa2. 3\spb3. 1^3}
+ \frac{\spab4.(1+2).3^2\spb2. 1^2}{2s_{123}\spb3. 1^3\spb3. 2}
- \frac{\spa1. 4^2\spb2. 1}{2\spa1. 2\spb3. 1\spb3. 2} \nn \\
+ \frac{\spb2. 1 \left( \spa1. 3^2\spa2. 3\spab4.(1+2).3^2\spb3. 1^2 + 
    \spa1. 2^3\spab4.(2+3).1^2\spb2. 1\spb3. 2 \right)}{
  4s_{123}^2\spa1. 2\spa2. 3\spb3. 1^3\spb3. 2} \nn \\ 
+ \frac{ \spab3.(1+4). 2^2}{2s_{124}\spb4. 1\spb4. 2}
- \frac{\spa1. 3^2\spb2. 1}{2\spa1. 2\spb4. 1\spb4. 2} 
+ \frac{\spa1. 4^2\spab3.(1+2).4^2\spb2. 1}{4s_{124}^2\spa1. 2\spb4. 1\spb4. 2} \nn \\
- \frac{\spa1. 3\spa1. 4\spb4. 2}{2\spab1.(2+3). 4\spb4. 3}
- \frac{s_{234}\spa1. 4^2\spb4. 2^2}{4\spa2. 3\spab1.(2+3). 4\spb3. 2^2\spb4. 3}
- \frac{\spa1. 4^2\spb4. 2^2}{2\spab1.(2+3). 4\spb3. 2\spb4. 3} \;.
\eea

\subsubsection{$A_4^{\fl}(\phi,1_\qb^-,2_q^+,\g3^-,\g4^-)$}
The fermion loop contribution is,
\bea
&&-i A_4^{\fl}(\phi,1_\qb^-,2_q^+,\g3^-,\g4^-)\ =\
-i A_4^\treenum(\phi,1_\qb^-,2_q^+,\g3^-,\g4^-)
\times\left[ - \frac{2}{3\e} \left(\frac{\mu^2}{-s_{12}}\right)^\e
- \frac{10}{9} \right] \nn \\
&&\null
       + \frac{2}{3} {\spa1.3}^2 \spa3.4 \spb1.2 \spab4.(1+2).3 \; \wh{\Ll}_3(s_{123},s_{12})
       + \frac{2}{3} {\spa1.4}^2 \spa3.4 \spb1.2 \spab3.(1+2).4 \; \wh{\Ll}_3(s_{124},s_{12})\nn \\
&&\null
       + \frac{1}{3} \spa1.3 \spa3.4 \spab4.(1+3).2 \; \wh{\Ll}_2(s_{123},s_{12})
        +\frac{1}{3} \spa1.4 \spa3.4 \spab3.(1+4).2 \; \wh{\Ll}_2(s_{124},s_{12})\nn \\
&&\null
        -\frac{2}{3} \frac{\spa1.3 {\spab4.(1+3).2}^2}
         {\spa1.2 \spb2.3 \spb1.2} \; \wh{\Ll}_1(s_{123},s_{12})
        +\frac{2}{3} \frac{\spa1.4 {\spab3.(1+4).2}^2}
         {\spa1.2 \spb2.4 \spb1.2} \; \wh{\Ll}_1(s_{124},s_{12})\nn \\
&&\null
       + \frac{2}{3} \frac{\spa1.3 {\spab4.(1+3).2}^2}
             {\spa1.2 \spb1.2 \spb2.3 s_{123}} \; \wh{\Ll}_0(s_{123},s_{12})
        +\frac{2}{3} \frac{(s_{12}+s_{14}) {\spab3.(1+4).2}^2}
           {\spa1.2 \spb1.4 \spb2.4 \spb1.2 s_{124}} \; \wh{\Ll}_0(s_{124},s_{12})\nn \\
&&\null
        -\frac{ \spa1.3 \spa3.4 \spab4.(1+3).2}{6 \spa1.2 \spb1.2 s_{123}}
        -\frac{ \spa1.4 \spa3.4 \spab3.(1+4).2}{6 \spa1.2 \spb1.2 s_{124}} 
          -\frac{1}{3} \frac{\spa1.3 \spa1.4 }{\spa1.2 \spb3.4} \;.
\label{qqgg_mpmm_fl} \\ \nn
\eea

\subsubsection{Relation for rational terms}
  
We note that the rational terms in the three leading colour primitive amplitudes obey,
\bea
&& {\cal R} \Big\{
 A_4^L(\phi,1_\qb,2_q,\g3,\g4)+A_4^R(\phi,1_\qb,2_q,\g3,\g4)
+A_4^{\fl}(\phi,1_\qb,2_q,\g3,\g4)\Big\} \nn \\
 && \qquad +2 A_4^\treenum(\phid,1_\qb,2_q,\g3,\g4)=0 \;,
\label{eq:ratrel}
\eea
a formula analogous to that found
in super-symmetric decompositions of QCD amplitudes~\cite{Bern:1994fz}. This property is helicity independent and has
also been checked for the previously known MHV amplitudes~\cite{Dixon:2009uk}.
For the NMHV helicity assignment at hand, namely $(1_\qb^-,2_q^+,\g3^-,\g4^-)$, we
note that the tree graph result that appears in Eq.~(\ref{eq:ratrel}) is zero (c.f. Eq.~(\ref{phidaqggmpmmtree})).
We stress that the ${\cal R}$ operation extracts the full rational 
term, including completion terms from the functions $\wh{\Ll}_3$ and $\wh{\Ll}_2$.
Thus it corresponds to dropping all logarithms, box functions and $V$-functions.  

We conclude this section by noting that the three primitive amplitudes 
for the helicity assignment $(1_{\bar{q}}^-,2_{q}^+,\g3^+,\g4^+)$ displayed in
Eqs.~(\ref{qqgg_mppp_L}),~(\ref{qqgg_mppp_R}) and~(\ref{qqgg_mppp_fl}), also satisfy Eq.~(\ref{eq:ratrel}).
For these amplitudes, which are purely rational, the ${\cal R}$ operation leaves the amplitude unchanged.

\subsection{Results for $A_{4;3}(\phi,1_\qb,2_q,\g3^-,\g4^-)$}
We can calculate the result for $A_{4;3}$ using Eq.~(\ref{A43defn}).
Given the results for $A^{4}_{L}$ and $A^{4}_{R}$ in the previous section 
the only missing ingredient is $A_4^L(\phi,1_\qb^-,\g2^-,3_q^+,\g4^-)$.

\subsubsection{Box-related terms for $A_4^L(\phi,1_\qb^-,\g2^-,3_q^+,\g4^-)$}
The calculation of the box-related terms in $\phi\qb gqg$ $({-}{-}{+}{-})$ 
is easily performed using the methods given in ref.~\cite{Britto:2004nc}. The result is,
\bea
&&\null -i A_4^{L,\mathrm{box}}(\phi,1_\qb^-,\g2^-,3_q^+,\g4^-)
\ =\
-i A_4^\treenum(\phi,1_\qb^-,\g2^-,3_q^+,\g4^-) \times V_4^L  \nn \\
&& \null
  +\frac{{\spa1.2}^2 m_{\phi}^4}
    {\spab1.(2+3).4 \spab3.(1+2).4 s_{123}}
    \Ls_{-1}(s_{12},s_{23};s_{123}) 
  -\frac{{\spab2.(1+4).3}^2}
    {\spb1.4  \spb3.4 s_{134}}
    \Ls_{-1}(s_{14},s_{34};s_{134})  \nn \\
&& \null
  +\frac{{\spb3.4}^2 \spab1.(3+4).2^2}
    {\spb3.2 {\spb2.4}^3  \spab1.(2+3).4}
    \Ls_{-1}(s_{23},s_{34};s_{234}) 
   +\frac{s_{124}^2}
     {\spb1.2 \spb2.4 \spab3.(1+2).4}
     \Ls_{-1}(s_{12},s_{14};s_{124})  \nn \\
&& \null
   +\frac{\spab3.(2+4).1 s_{124}^2}
    {\spb1.2 \spb1.4 \spab3.(1+2).4 \spab3.(1+4).2}
    \Lsnew^{2{\rm m}h}_{-1}(s_{23},s_{124};s_{14},m_\phi^2)  \nn \\
&& \null
   -\frac{\spab1.(2+4).3^3}
    {\spb2.3  \spb3.4  \spab1.(2+3).4  \spab1.(3+4).2}
    \Lsnew^{2{\rm m}h}_{-1}(s_{12},s_{234};s_{34},m_\phi^2)  \nn \\
&& \null
 +\frac{1}{s_{123}} \Bigg[\frac{m_{\phi}^4 {\spa1.2}^2}
   {\spab3.(1+2).4   \spab1.(2+3).4 }
  +\frac{{\spab4.(1+2).3}^2}
   {\spb1.2 \spb2.3 } \Bigg] \nn \\
&& \null
   \times \Bigg[ \Lsnew^{2{\rm m}h}_{-1}(s_{34},s_{123};s_{12},m_\phi^2) 
                +\Lsnew^{2{\rm m}h}_{-1}(s_{14},s_{123};s_{23},m_\phi^2) \Bigg] \;,
\eea
with
\bea
V_4^L &=& - \frac{1}{\e^2} \left[
\left(\frac{\mu^2}{-s_{34}}\right)^\e +
\left(\frac{\mu^2}{-s_{41}}\right)^\e \right] +
\frac{1}{3\e}\left(\frac{\mu^2}{-s_{123}}\right)^\e +
\frac{7}{4} - \frac{\delta_R}{3} \;.
\eea
As we shall see in the next section, no further information is required 
for the calculation of the $A_{4;3}$ which is completely determined 
by box diagrams alone.

\subsubsection{Full result for $A_{4;3}$}
The full result for the partial amplitude $A_{4;3}$ is,
\bea
&-&i A_{4;3}(\phi,1_\qb^-,2_q^+,\g3^-,\g4^-) = 
-i A_4^\treenum(\phi,1_\qb^-,2_q^+,\g3^-,\g4^-) \times V_5(s_{12}, s_{34}, s_{13}, s_{24})
 \nn \\ &&\null
  +\frac{1}{s_{123}} \left[ \frac{\spab4.(1+2).3^2 \spb1.2^2}{\spb1.3^3 \spb2.3}
  + \frac{\spb2.3^2 \spab4.(2+3).1^3}
    {\spb1.3^3 \spb1.2 \spab4.(1+2).3}
  -\frac{m_\phi^4 \spa1.3^3}
   {\spa1.2 \spab1.(2+3).4 \spab3.(1+2).4} \right] \nn \\
   &&\null \quad \times
   \Ls_{-1}(s_{12},s_{23};s_{123})
 \nn \\ &&\null
  +\frac{1}{s_{124}} \left[ \frac{m_\phi^4 \spa1.4^2 \spa2.4}
   {\spa1.2 \spab2.(1+4).3 \spab4.(1+2).3}
  -\frac{\spab3.(1+4).2^2 \spab3.(2+4).1}
   {\spb1.4 \spab3.(1+2).4 \spb1.2} \right]
   \Ls_{-1}(s_{12},s_{14};s_{124})
 \nn \\ &&\null
  +\frac{1}{s_{123}} \left[ \frac{m_\phi^4 \spa1.3^2}{\spab1.(2+3).4 \spab2.(1+3).4}
  -\frac{\spab4.(1+3).2^2}{\spb1.3 \spb2.3} \right]
   \Ls_{-1}(s_{13},s_{23};s_{123})
 \nn \\ &&\null
  +\frac{s_{341}^2}{\spb1.3 \spb3.4 \spab2.(1+3).4}
   \left[ \Ls_{-1}(s_{13},s_{14};s_{341})
         +\Lsnew^{2{\rm m}h}_{-1}(s_{23},s_{341},s_{14},m_\phi^2) \right]
 \nn \\ &&\null
  +\frac{s_{341}^2}{\spb1.4 \spb3.4 \spab2.(1+4).3}
   \left[ \Ls_{-1}(s_{14},s_{34};s_{341})
         +\Lsnew^{2{\rm m}h}_{-1}(s_{12},s_{341},s_{34},m_\phi^2) \right]
 \nn \\ &&\null
  -\frac{\spab1.(3+4).2^2}{\spb2.3 \spb3.4 \spab1.(2+3).4}
   \left[ \Ls_{-1}(s_{23},s_{34};s_{234})
         +\Lsnew^{2{\rm m}h}_{-1}(s_{12},s_{234},s_{34},m_\phi^2) \right]
 \nn \\ &&\null
  +\frac{\spb2.4^2 \spab1.(2+4).3^2}{\spb2.3 \spb3.4^3 \spab1.(2+3).4}
   \Ls_{-1}(s_{23},s_{24};s_{234})      
 \nn \\ &&\null
  -\frac{\spab1.(3+4).2^2}{\spab1.(2+4).3 \spb2.4 \spb3.4}
   \Lsnew^{2{\rm m}h}_{-1}(s_{14},s_{234},s_{23},m_\phi^2) 
 \nn \\ &&\null
  +\frac{1}{s_{123}} \left[ -\frac{m_\phi^4 \spa1.3^2 \spa2.3}
   {\spa1.2 \spab2.(1+3).4 \spab3.(1+2).4}
  +\frac{\spab4.(1+3).2^2}{\spab4.(1+2).3}
   \frac{\spab4.(2+3).1}{\spb1.2 \spb1.3} \right]
 \nn \\ &&\null
 \qquad \times \Lsnew^{2{\rm m}h}_{-1}(s_{14},s_{123},s_{23},m_\phi^2)
 \nn \\ &&\null
  +\frac{1}{s_{123}} \left[ \frac{m_\phi^4 \spa1.3^3}{\spa1.2
   \spab3.(1+2).4 \spab1.(2+3).4}
  -\frac{\spab4.(1+3).2^3}{\spab4.(1+2).3 \spb1.2 \spb2.3} \right]
 \nn \\ &&\null
 \qquad \times \Lsnew^{2{\rm m}h}_{-1}(s_{24},s_{123},s_{13},m_\phi^2)
 \nn \\ &&\null
  +\frac{1}{s_{123}} \left[ -\frac{m_\phi^4 \spa1.3^2}
   {\spab2.(1+3).4 \spab1.(2+3).4}
  +\frac{\spab4.(1+3).2^2}{\spb1.3 \spb2.3} \right]
   \Lsnew^{2{\rm m}h}_{-1}(s_{34},s_{123},s_{12},m_\phi^2)
   \nn \\
&&\null
+ \Bigg\{3 \leftrightarrow 4\Bigg\} \;,
\eea
where the function containing poles and associated logarithms is conveniently written as,
\be
V_5(s_{12}, s_{34}, s_{13}, s_{24}) = - \frac{1}{\e^2} \left[
  \left(\frac{\mu^2}{-s_{12}}\right)^\e
+ \left(\frac{\mu^2}{-s_{34}}\right)^\e
- \left(\frac{\mu^2}{-s_{13}}\right)^\e
- \left(\frac{\mu^2}{-s_{24}}\right)^\e
 \right] \;.
\label{V5def}
\ee
We note that the apparent double pole in $\e$ in Eq.~({\ref{V5def}}) is cancelled
upon expanding about $\e = 0$.

This result for the $\phi$ amplitude is particularly simple,
containing neither bubble contributions nor rational terms. This 
is also true for the helicity amplitude $A_{4;3}(\phi,1_\qb,2_q,\g3^-,\g4^+)$~\footnote{
The amplitude $A_{4;3}(\phi,1_\qb,2_q,\g3^+,\g4^-)$ is not independent and is
obtained by swapping labels $3$ and $4$.},
which can easily be checked using the previously calculated results in ref.~\cite{Dixon:2009uk}. 
It is therefore more efficient to program the full result
for $A_{4;3}$, rather than to program the individual primitive amplitudes using 
Eq.~(\ref{A43defn}).

Furthermore, for the case of two negative gluon helicities calculated here 
one can check using Eq.~(\ref{A43defn})
and Eqs.~(\ref{phidaqgg_mpmm_L}),~(\ref{phidaqgg_mpmm_R}),~(\ref{phig_qgqg_mmpm_L})
that the corresponding $\phid$ amplitude is zero. Therefore we have,
\be
A_{4;3}(H,1_\qb^-,2_q^+,\g3^-,\g4^-)
 = i A_{4;3}(A,1_\qb^-,2_q^+,\g3^-,\g4^-)
 = A_{4;3}(\phi,1_\qb^-,2_q^+,\g3^-,\g4^-) \;.
\ee

\section{Numerical results \label{sec:numerical}}
Here we present evaluations of the new amplitudes at the same kinematic
point as used previously in the literature~\cite{Ellis:2005qe,Dixon:2009uk}:
\bea
k_\phi &=& (-1.0000000000,\ \, 0.00000000000,\ \ 0.00000000000,\ \ 0.00000000000),\nn\\
k_1 &=& (0.30674037867,-0.17738694693,-0.01664472021,-0.24969277974),\nn\\
k_2 &=& (0.34445032281,\ \ 0.14635282800,-0.10707762397,\ \ 0.29285022975),
~~~\label{EGZDSpoint}\\
k_3 &=& (0.22091667641,\ \ 0.08911915938,\ \ 0.19733901856,\ \ 0.04380941793),\nn\\
k_4 &=& (0.12789262211,-0.05808504045,-0.07361667438,-0.08696686795).\nn
\eea
We have used a scale $\mu = m_H$, set $\delta_R = 1$ (corresponding to the 't Hooft-Veltman scheme)
and, in assembling the amplitude $A_{4;1}$, have used $n_f = 5$. The results for the final
Higgs amplitudes presented in Table~\ref{numres} agree with those from the semi-numerical calculation
of ref.~\cite{Ellis:2005qe} to one part in $10^8$.
Note that these results depend on an overall phase that can be removed by
dividing out by the corresponding Born calculation.
Using the analytic expressions for all the $Hgggg$, $H\bar{q}qgg$ and
$H\bar{q}q \bar{q}^\prime q^\prime$ amplitudes that are now available
we can also confirm\footnote{
Fortran code that calculates all the amplitudes can be downloaded from {\tt mcfm.fnal.gov}.}
the numerical values for the matrix elements squared given in ref.~\cite{Ellis:2005qe}.
\TABLE[t]{ \begin{tabular}{|c||c|c|c||c|}
\hline
& \multicolumn{3}{c||}{$(\phi,1_\qb^-,2_q^+,\g3^-,\g4^-)$} & $(\phid,1_\qb^-,2_q^+,\g3^-,\g4^-)$ \\
\hline  & $1/\e^2$ & $1/\e$ & $\e^0$ & $\e^0$ \\
\hline
\hline $A_4^{(0)}$
& $           0          $ & $           0          $ & $     +6.49907535901   $ & $           0          $ \\ 
& $                     $ & $                       $ & $     -2.39308144816\,i$ & $                      $ \\ 
\hline $A_4^L$
& $    -19.49722607702   $ & $    -64.62496304875   $ & $    -31.60558356648   $ & $    -17.35549203005   $ \\ 
& $     +7.17924434447\,i$ & $    -45.76112071571\,i$ & $   -137.56039301452\,i$ & $     +6.14361664194\,i$ \\ 
\hline $A_4^R$
& $     -6.49907535901   $ & $    -23.12631834140   $ & $    -48.74190400225   $ & $     +4.58546771410   $ \\ 
& $     +2.39308144816\,i$ & $    -14.67020390044\,i$ & $    -39.06265552875\,i$ & $     -1.38718545292\,i$ \\ 
\hline $A_4^f$
& $           0          $ & $     -4.33271690600   $ & $    -14.98058321393   $ & $     -0.22812640212   $ \\ 
& $                      $ & $     +1.59538763210\,i$ & $     -9.88874973495\,i$ & $     +0.02973170727\,i$ \\ 
\hline
\hline $A_{4;1}$
& $    -18.77510659269   $ & $    -69.27656696526   $ & $    -51.15745514499   $ & $    -18.24519911293   $ \\ 
& $     +6.91334640579\,i$ & $    -41.47211867326\,i$ & $   -149.70134751402\,i$ & $     +6.34730120438\,i$ \\ 
\hline $A_{4;3}$
& $           0          $ & $     +2.61083477136   $ & $    +17.75737443413   $ & $           0          $ \\ 
& $                      $ & $     -0.05119106396\,i$ & $     +4.93097014463\,i$ & $                      $ \\ 
\hline
\end{tabular}
\caption{Numerical values of $\phi\qb qgg$ and $\phid\qb qgg$ primitive amplitudes (above)
and the amplitudes multiplying the two different colour structures (below),
at the kinematic point defined in Eq.~(\ref{EGZDSpoint}).
\label{numres} } }

\section{Conclusions \label{sec:conclusions}}

In this paper we have computed the last remaining, analytically
unknown, helicity amplitude contributing to the NLO corrections to
Higgs plus two jet production at hadron colliders. This builds upon
the previously known semi-numerical results~\cite{Ellis:2005qe} and
completes the set of compact analytic
formulae~\cite{Badger:2006us,Berger:2006sh,Badger:2007si,Glover:2008ffa,Dixon:2009uk,Badger:2009hw}.

We employed a generalised unitarity approach to calculate the cut-constructible
parts of the amplitude. Completion of the logarithmic terms to remove
unphysical singularities was used to simplify rational terms extracted
from a Feynman diagram calculation. Simplifications in the
construction of the subleading colour amplitude, $A_{4;3}$, showed
that this component is free from all rational, bubble and triangle
terms. Similar relations between the rational terms in
the left, right and fermion loop primitive amplitudes were used to
find a compact analytic structure.

Our results have been verified against the known numerical results and we envisage that they will provide the means for a
faster and more flexible analysis of Higgs phenomenology at the Tevatron and the LHC.

\acknowledgments
We are happy to acknowledge useful discussions with Nigel Glover,
Pierpaolo Mastrolia and Giulia Zanderighi.
CW acknowledges the award of an STFC studentship and
SB acknowledges support from
the Helmholtz Gemeinschaft under contract VH-NG-105.
Fermilab is operated by Fermi Research Alliance, LLC under
Contract No. DE-AC02-07CH11359 with the United States Department of Energy.

\appendix 
\section{Spinor notation \label{spinnotation}}
Our spinor notation is quite standard in the 
QCD literature, (for a review see refs.~\cite{Mangano:1990by,Dixon:1996wi})
The function $u_\pm(k_i)$ is a
massless Weyl spinor of momentum $k_i$ and positive or negative
chirality. In terms of these solutions of the Dirac equation, the spinor products are 
defined by,
\bea
\spa{i}.{j} &=& 
             = \langle i^-|j^+\rangle = \bar{u}_-(k_i)u_+(k_j)\,,
             \label{defspa}\\
\spb{i}.{j} &=& 
             = \langle i^+|j^-\rangle = \bar{u}_+(k_i) u_-(k_j)\,.
             \label{defspb}
\eea
We use the convention $\spb{i}.{j} = \mathop{\rm
sgn}(k_i^0k_j^0)\spa{j}.{i}^*$, so that,
\be
\spa{i}.{j}\spb{j}.{i} = 2k_i\cdot k_j \equiv s_{ij} \,.
\ee
We further define,
\be
s_{ijl} \equiv (k_i+k_j+k_l)^2
= \spa{i}.{j}\spb{j}.{i} + \spa{j}.{l} \spb{l}.{j}
+ \spa{i}.{l} \spb{l}.{i} \;,
\ee
\be
\langle a|i|b]\ =\ \spa{a}.{i}\spb{i}.{b} \,, \qquad\quad
\langle a|(i+j)|b]
\ =\ \spa{a}.{i}\spb{i}.{b} + \spa{a}.{j}\spb{j}.{b} \,,
\label{spinorsandwich}
\ee
\be
\spa{j}.{i} = - \spa{i}.{j} \,, \qquad\quad \spb{j}.{i} = - \spb{i}.{j} \,.
\ee
Simplification of the formula can sometimes be achieved by using 
the Schouten identity,
\bea
\spa{a}.{b}\spa{c}.{d} &=&
\spa{a}.{d}\spa{c}.{b} + \spa{a}.{c}\spa{b}.{d} \,, \\
\spb{a}.{b}\spb{c}.{d} &=&
\spb{a}.{d}\spb{c}.{b} + \spb{a}.{c}\spb{b}.{d} \,.
\eea

\section{Definitions of special functions \label{intnotation}}

With the definition of the $\Ll_i(s,t)$ basis functions, 
\begin{eqnarray}
\Ll_i(s,t)=\frac{\log{(s/t)}}{(s-t)^i} \;,
\end{eqnarray}
we can define the completions,
\begin{eqnarray}
\Ll_3(s,t) &\to& \wh{\Ll}_3(s,t) = \Ll_3(s,t)-\frac{1}{2(s-t)^2} \left(\frac{1}{s}+\frac{1}{t}\right) \;, \nn \\
\Ll_2(s,t) &\to& \wh{\Ll}_2(s,t) = \Ll_2(s,t)-\frac{1}{2(s-t)} \left(\frac{1}{s}+\frac{1}{t}\right) \;, \nn \\
\Ll_1(s,t) &\to& \wh{\Ll}_1(s,t) = \Ll_1(s,t) \;,
\end{eqnarray}
such that the completed functions are finite in the limit that $s \to t$.
Note that the definition of these completed logarithmic functions is similar in spirit,
but different in detail from the definitions in ref.~\cite{Dixon:2009uk}.  

We also need the box functions from the scalar box with one massive external leg,
\begin{eqnarray}
\Ls_{-1}(s,t;m^2) &=&\Li_2\left(1-\frac{s}{m^2}\right)+\Li_2\left(1-\frac{t}{m^2}\right)
+\log\left(\frac{s}{m^2}\right) \log\left(\frac{t}{m^2}\right)-\frac{\pi^2}{6} \;,
\end{eqnarray}
and coming from the box with two adjacent massive external legs,
\begin{eqnarray}
\Lsnew^{2{\rm m}h}_{-1}(s,t;m_1^2,m_2^2) &=&
    -\Li_2\left(1-{m_1^2\over t}\right)
    -\Li_2\left(1-{m_2^2\over t}\right)
    -\frac{1}{2}\log^2\left({-s\over-t}\right) \nn \\
    &+&\frac{1}{2}\log\left({-s\over-m_1^2}\right)
              \log\left({-s\over-m_2^2}\right)\,.
\end{eqnarray}
where the dilogarithm is defined as usual by,
\begin{equation}
\Li_2(x) = - \int_0^x dy \; {\log(1-y) \over y}\,.
\end{equation}
$\I33m$ is the three mass triangle function defined, for example,
in Eq.~(II.9) of ref.~\cite{Bern:1997sc},
\be
\I33m(s_{12},s_{34},s_{56})\ =\ 
\int_0^1 d^3a_i\, \delta(1-a_1-a_2-a_3)\, 
\frac{1} {[-s_{12}a_1a_2-s_{34}a_2a_3-s_{56}a_3a_1]}\; .
\ee
Explicit results for this integral can be found in refs.~\cite{Lu:1992ny},
\cite{Davydychev:1995mq} and~\cite{Usyukina:1993ch}.

\end{document}